\documentclass[%
 reprint,
 amsmath,
 amssymb,
 aps,
 prx,
]{revtex4-1}

\usepackage{graphicx}
\usepackage{dcolumn}
\usepackage{bm}
\usepackage{epsfig,color,xspace,multirow,xr,bbold}
\usepackage[all]{xy}
\usepackage{setspace}
\usepackage{url}
\usepackage{siunitx}

\newcommand{\dGwi}{\Delta G_{\textup{W}\rightarrow\textup{I}}}
\newcommand{\dGim}{\Delta G_{\textup{I}\rightarrow\textup{M}}}
\newcommand{\dGwm}{\Delta G_{\textup{W}\rightarrow\textup{M}}}
\newcommand{\dGwol}{\Delta G_{\textup{W}\rightarrow\textup{Ol}}}

\usepackage{hyperref} 

\hypersetup{
   colorlinks,
   menucolor=blue,
   linkcolor=black,
   citecolor=blue,
   urlcolor=blue
}

\begin{document}

\title{Controlled exploration of chemical space by machine learning of
  coarse-grained representations}

\author{Christian Hoffmann}
\author{Roberto Menichetti}
\author{Kiran H.~Kanekal}
\author{Tristan Bereau}
\email{bereau@mpip-mainz.mpg.de}

\affiliation{Max Planck Institute for Polymer Research, 55128 Mainz,
  Germany}

\date{\today}

\begin{abstract}
  The size of chemical compound space is too large to be probed
  exhaustively.  This leads high-throughput protocols to drastically
  subsample and results in sparse and non-uniform datasets.  Rather
  than arbitrarily selecting compounds, we systematically explore
  chemical space according to the target property of interest.  We
  first perform importance sampling by introducing a Markov chain
  Monte Carlo scheme across compounds.  We then train a machine
  learning (ML) model on the sampled data to expand the region of chemical 
  space probed.  Our boosting procedure enhances the number of compounds
  by a factor 2 to 10, enabled by the ML model's coarse-grained representation, 
  which both simplifies the structure-property relationship and reduces the
  size of chemical space.  The ML model correctly recovers linear
  relationships between transfer free energies.  These linear
  relationships correspond to features that are global to the dataset,
  marking the region of chemical space up to which predictions are
  reliable---a more robust alternative to the predictive variance.
  Bridging coarse-grained simulations with ML gives rise to an
  unprecedented database of drug-membrane insertion free energies for
  1.3 million compounds.
\end{abstract}

\maketitle

\section{Introduction}

Computational high-throughput screening ever-increasingly allows the
coverage of larger subsets of chemical space.  Extracting a property
of interest across many compounds helps infer structure-property
relationships, of interest both for a better understanding of the
physics and chemistry at hand, as well as for materials
design~\cite{pyzer2015high, jain2016computational, Bereau2016,
von2018quantum}.  Recent hardware and algorithmic developments have
enabled a number of applications of high-throughput screening in hard
condensed matter~\cite{curtarolo2013high, ghiringhelli2015big}, while
comparatively slower development in soft
matter~\cite{ferguson2017machine, bereau2018data}.

Soft-matter systems hinge on a delicate balance between enthalpic and
entropic contributions, requiring proper computational methods to
reproduce them faithfully.  Physics-based methodologies, in particular
molecular dynamics (MD), provide the means to systematically sample
the conformational ensemble of complex systems.  High-throughput
screening using MD presumably involves one simulation per compound.
This remains computationally prohibitive at an atomistic resolution.

As an alternative, we recently proposed the use of coarse-grained (CG)
models to establish a high-throughput scheme.  Coarse-grained models
lump several atoms into one bead to decrease the number of degrees of
freedom~\cite{Noid2013}.  The computational benefit of a
high-throughput coarse-grained (HTCG) framework is two-fold: ($i$)
faster sampling of conformational space; but most importantly ($ii$) a
significant reduction in the size of chemical space, tied to the
transferable nature of the CG model (i.e., a finite set of bead
types).  Effectively the reduction in chemical space due to
coarse-graining still leads to a combinatorial explosion of chemical
space, but with a significantly smaller prefactor.  This many-to-one
mapping is empirically probed by coarse-graining large databases of
small molecules---an effort made possible by automated force-field
parametrization schemes~\cite{Bereau2015}.  Using the CG Martini force
field~\cite{periole2013martini}, we recently reported the
\emph{exhaustive} characterization of CG compounds made of one and two
beads, corresponding to small organic molecules between 30 and 160~Da.
Running HTCG for 119 CG compounds enabled the predictions of
drug-membrane thermodynamics~\cite{menichetti2017silico} and
permeability~\cite{menichetti2018drug} for more than $500,000$ small
organic molecules. 

Pushing further our exploration of chemical space, how do we further
diversify the probed chemistry in a systematic manner?  Scaling up to
larger molecular weight using more CG beads will ultimately lead to a
combinatorial explosion: there are already 1,470 linear trimers and
19,306 linear tetramers in Martini. Instead of an exhaustive
enumeration, we propose to explore regions of chemical space that are
of particular interest. Specifically, rather than following the
chemistry we navigate chemical space according to a \emph{target
property}---in this work, the tendency of a small organic molecule to
partition at the interface of a phospholipid bilayer.  

\begin{figure*}[htbp]
  \begin{center}
    \includegraphics[width=\linewidth]{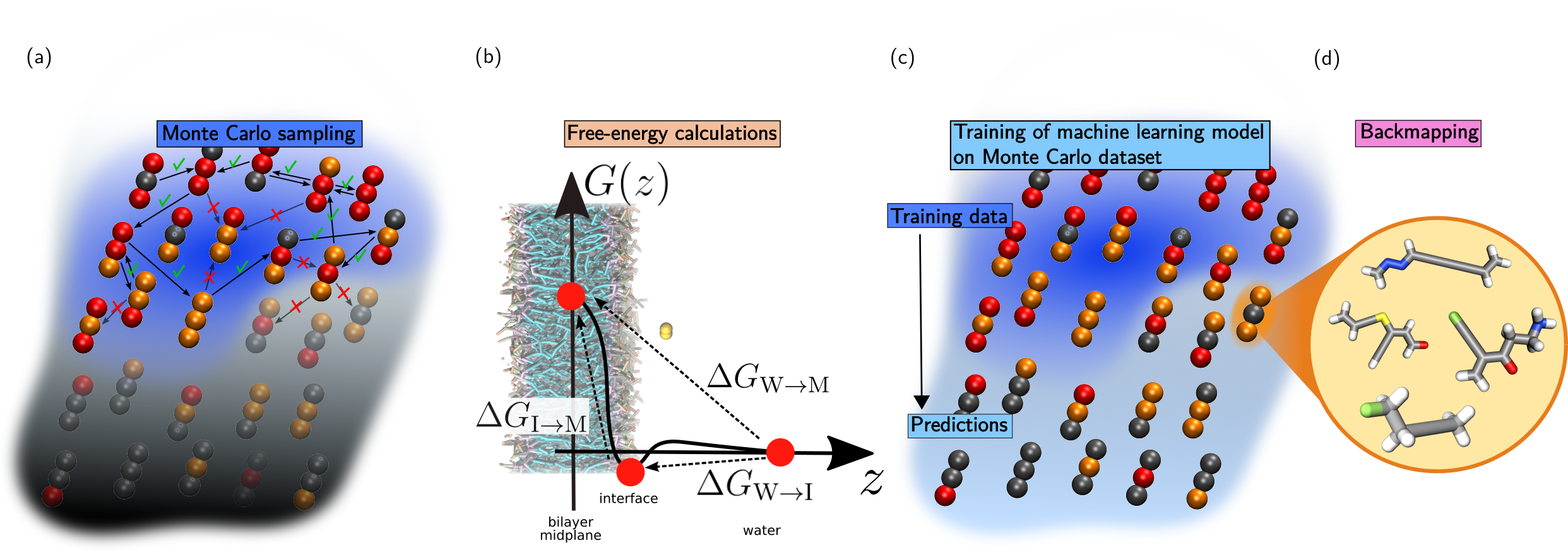}
    \caption{(a) Importance sampling across coarse-grained compounds
      via a Markov chain Monte Carlo scheme.  Only the dark-blue
      region is sampled.  (b) Background: Simulation setup of a solute
      (yellow) partitioning between water (not shown) and the lipid
      membrane.  Foreground: Potential of mean force along the normal
      of the bilayer, $G(z)$, and definition of the three transfer
      free energies of interest between the three state points (red
      circles): bilayer midplane (``M''), membrane-water interface
      (``I''), and bulk water (``W'').  (c) The MC-sampled free
      energies (dark-blue region) form the training set for a machine
      learning model, used to predict a larger subset of compounds
      (light-blue region).  (d) Each coarse-grained compound
      represents a large number of small molecules.}
    \label{fig:intro}
  \end{center}
\end{figure*}

Akin to importance sampling used extensively to characterize
conformational space, we first introduce a Markov chain Monte Carlo
(MC) procedure to sample chemical compound space.  Our methodology
consists of a sequence of compounds where trial alchemical
transformations are accepted according to a Metropolis criterion
(Fig.~\ref{fig:intro}a).  While the acceptance criterion for
conformational sampling typically includes the energy of the system,
compositional sampling (i.e., across chemical space) means averaging
over the environment and thus calls for a free energy.   As such, MC
moves will push the exploration towards small molecules that increase
their stability within a specific condensed-phase
environment---algorithmically akin to constant pH
simulations~\cite{mongan2004constant}.

In this work we focus on the free-energy difference of transferring a
small molecule from the aqueous environment to the lipid-membrane
interface (Fig.~\ref{fig:intro}b). The MC acceptance criterion is here
dictated by pharmacokinetic considerations: while hydrophobic
compounds will more easily permeate through the bilayer
\cite{menichetti2018drug}, they may display poor solubility
properties~\cite{dahan2016solubility}. Our MC criterion thus aims at
balancing the delicate interplay between solubility and permeability.

To further boost the size of chemical space that is probed, we further
predict a more extended subset of compounds that were not sampled by
using machine learning (ML; see
Fig.~\ref{fig:intro}c)~\cite{rasmussen2004gaussian}.  Despite known
limited capabilities to extrapolate beyond the training set, we
observe remarkable accuracy for the predicted compounds.  This
excellent transferability can be associated to a simplified learning
procedure at the CG resolution: structure-property relationships are
easier to establish~\cite{menichetti2018drug} and compound similarity
is compressed due to the reduction of chemical space.  The range of
reliable predictions is made clear by means of the ML model satisfying
linear thermodynamic relations across
compounds~\cite{menichetti2017silico}---a more robust confidence
metric compared to the predictive variance.  The CG results are then
systematically backmapped (Fig.~\ref{fig:intro}d) to yield an
unprecedentedly-large database of free energies.

\section{Methods}

\subsection{Coarse-grained simulations}
MD simulations of the Martini force field \cite{periole2013martini}
were performed in {\sc Gromacs} 5.1. The integration time-step was
$\delta t=0.02~\tau$, where $\tau$ is the model's natural unit of
time.  Control over the system temperature and pressure ($T=300~K$ and
$P=1$~bar) was obtained by means of a velocity rescaling thermostat
\cite{bussi2007canonical} and a Parrinello-Rahman barostat
\cite{parrinello1981polymorphic}, with coupling constants
$\tau_T=~\tau$ and $\tau_P=12~\tau$.  Bulk simulations consisted of
$N_{\textup{W}}=450$ and $N_{\textup{O}}=336$ water and octane
molecules, where the latter was employed as a proxy for the
hydrophobic core of the bilayer \cite{menichetti2017silico}.  As for
interfacial simulations, a membrane of $36~\text{nm}^2$ containing
$N_{\textup{L}}=128$ 1,2-dioleoyl-\emph{sn}-glycero-3-phosphocholine (DOPC) lipids (64 per layer) and
$N_{\textup{W}}'=1890$ water molecules was generated by means of the
{\sc Insane} building tool \cite{wassenaar2015computational}, and
subsequently minimized, heated up, and equilibrated.  In all
simulations containing water molecules we added an additional $10\%$
of antifreeze particles.

\subsection{Free-energy calculations}
\label{sec:free_en}

Water/interface and interface/membrane transfer free energies $\dGwi$
and $\dGim$ for all compounds investigated in this work were obtained
from alchemical transformations, in analogy with
Ref.~\cite{menichetti2017silico}.  This construction is based on the
relation linking the transfer free energies of two compounds $A$ and
$B$ ($\Delta G^{A}_{\textup{W}\rightarrow\textup{I}}$,
$\Delta G^{B}_{\textup{W}\rightarrow\textup{I}}$ and
$\Delta G^{A}_{\textup{I}\rightarrow\textup{M}}$,
$\Delta G^{B}_{\textup{I}\rightarrow\textup{M}}$) to the free energies
of alchemically transforming $A$ into $B$ in the three fixed
environments, $\Delta G^{A\rightarrow B}_{\textup{I}}$,
$\Delta G^{A\rightarrow B}_{\textup{W}}$ and
$\Delta G^{A\rightarrow B}_{\textup{M}}$

 \begin{equation}
\nonumber
  \Delta G^{B}_{\textup{W}\rightarrow\textup{I}}=\Delta
  G^{A}_{\textup{W}\rightarrow\textup{I}}+(\Delta G^{A\rightarrow
    B}_{\textup{I}}-\Delta G^{A\rightarrow B}_{\textup{W}}),
\end{equation}
\begin{equation}
\label{eq:alch_transf}
  \Delta G^{B}_{\textup{I}\rightarrow\textup{M}}=\Delta
  G^{A}_{\textup{I}\rightarrow\textup{M}}+(\Delta G^{A\rightarrow
    B}_{\textup{M}}-\Delta G^{A\rightarrow B}_{\textup{I}}).
\end{equation}

$\Delta G^{A\rightarrow B}_{\textup{I}}$, $\Delta G^{A\rightarrow
B}_{\textup{W}}$, and $\Delta G^{A\rightarrow B}_{\textup{M}}$ were
determined by means of separate MD simulations at the interface, in
bulk water, and in bulk octane.  For the calculation of each $\Delta
G^{A\rightarrow B}_i$, $i={\textup{I}},{\textup{W}},{\textup{M}}$ we
relied on the multistate Bennett acceptance ratio (MBAR)
\cite{shirts2008statistically, klimovich2015guidelines}, in which
free-energy differences are obtained by combining together the results
from simulations that sample the statistical ensemble of a set of
interpolating Hamiltonians $H(\lambda)$, $\lambda\in [0,1]$, with
$H(0)=H_A$ and $H(1)=H_B$.  We employed 24 evenly spaced
$\lambda$-values for each alchemical transformation and in each
environment (interface, water, octane). The production time for each
$\lambda$ point was $4 \cdot 10^{4}~\tau$ at the interface and $2
\cdot 10^{4}~\tau$ in bulk environments. To calculate $\Delta
G^{A\rightarrow B}_{\textup{I}}$ we added a harmonic potential with
$k=240~{\rm kcal}\,{\rm mol}^{-1}\,{\rm nm}^{-2}$ between the
compound's center of mass and the bilayer midplane at a distance
$\bar{z}=1.5~{\rm nm}$, to account for the spatial localization of the
interface. The value of $\bar{z}$ was fixed by analyzing the potential
of mean force $G(z)$ (see Fig.~\ref{fig:intro}b) for the insertion of
various solutes that preferentially sit near the lipid headgroups in a
DOPC bilayer~\cite{menichetti2017silico}. The minimum of these
profiles was found to be located at $\bar{z}\approx1.8~{\rm nm}$
irrespective of the compound's chemical detail, suggesting that the
location of the dip is largely determined by the membrane environment.
In this work, we corrected $\bar{z}$ to account for the horizontal
shift in the potentials of mean force generated by the additional bead
of the Martini DOPC model originally employed in
Ref.~\onlinecite{menichetti2017silico}~\cite{menichetti2017efficient}.

We further emphasize that we only restrained the compound's center of
mass, while the \emph{orientation} of the linear molecule with respect
to the bilayer normal was left unbiased. Notably, we do expect (and
observed) compounds to display very different preferential
orientations---from parallel to perpendicular with respect to the
bilayer normal. There are two reasons motivating our choice: ($i$) the
CG simulations efficiently explore conformational space anyway, such
that this degree of freedom is relatively easily sampled; and ($ii$)
the information between interpolating Hamiltonians that are simulated
during an alchemical transformation are efficiently exchanged thanks
to the MBAR method. The small corrections operated during the
thermodynamic-cycle optimization we apply a posteriori attest of our
assumptions.

\subsection{Monte Carlo sampling}

We perform a stochastic exploration of the chemical space of CG linear
trimers and tetramers through the generation of Markovian sequences of
compounds. Given the last compound $A$ of a sequence, the new compound
$B$ is proposed by randomly selecting a bead of $A$ and changing its
type. The move from $A$ to $B$ is then accepted with probability
\begin{equation}
  \label{eq:metr_mc}
  P_{A\rightarrow B} = \min \left\{ 1, \exp \left[-\beta(
  \Delta G^{B}_{\textup{W}\rightarrow\textup{I}} -
  \Delta G^{A}_{\textup{W} \rightarrow\textup{I}})\right] \right\}, 
\end{equation}
where $\Delta G^{A}_{\textup{W}\rightarrow\textup{I}}$ and $\Delta
G^{B}_{\textup{W}\rightarrow\textup{I}}$ are the water/interface
transfer free energies of $A$ and $B$, respectively. $P_{A\rightarrow
B}$ aims at driving the Monte Carlo sampling towards compounds that
favor partitioning at the membrane interface (Fig.~\ref{fig:intro}b).
This is significantly different from optimizing $\dGwm$, because those
would likely be poorly soluble in an aqueous
environment~\cite{dahan2016solubility}.

While in this work we set $\beta=1/k_\mathrm{B}T$, we stress
that $\beta$ can in principle be chosen independently of the system
temperature.  The free-energy difference in Eq.~\ref{eq:metr_mc} is
derived from the alchemical free-energy differences of transforming
$A$ into $B$ in the three fixed environments
$\Delta G^{A\rightarrow B}_i,~i=\textup{W, I, M}$ (first
relation in Eq.~\ref{eq:alch_transf}), which we compute from MD
simulations.
 
We generated up to five independent Markovian sequences in parallel,
each starting from a different initial compound. To avoid
recalculating alchemical transformations already visited, we stored
the history of calculations and looked up previously-calculated values
when available.

\subsection{Thermodynamic-cycle optimization}
\label{sec:topt}

\begin{figure}[htbp]
  \begin{center}
    \includegraphics[width=0.95\linewidth]{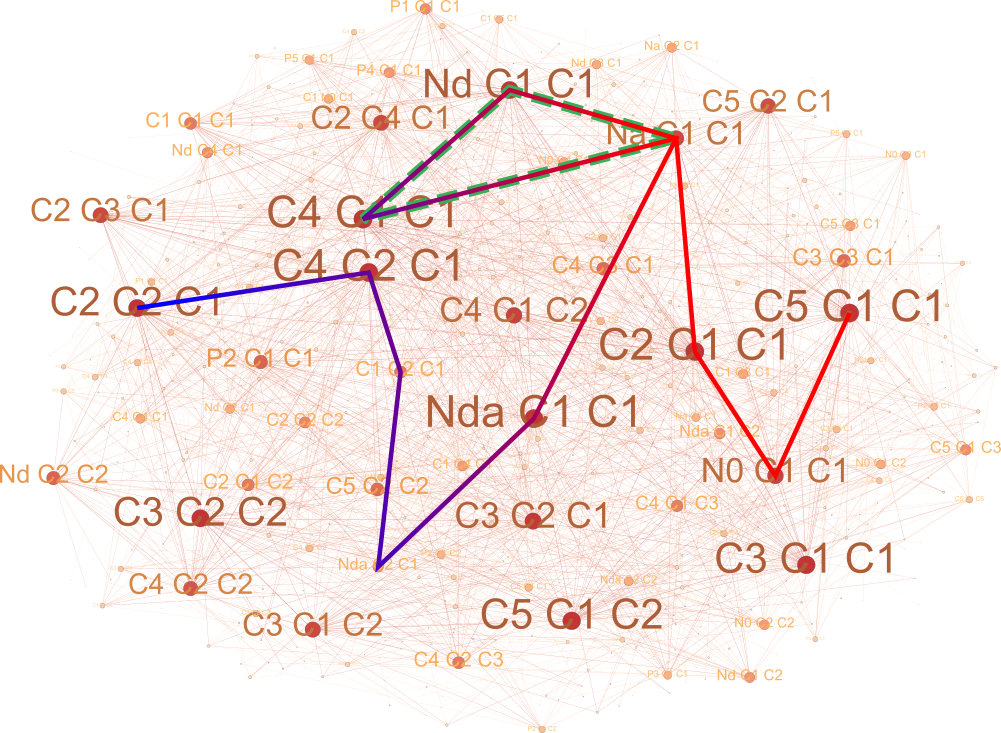}
    \caption{Network of CG compounds, each denoted by its set of
      Martini bead types. Their size is proportional to the number of
      edges. The blue-to-red path illustrates a sequence of accepted
      compounds sampled from the Monte Carlo scheme. The dashed, green
      triangle denotes a closed path in compound space. Sampling all
      of its three branches closes the thermodynamic cycle, used as
      constraint to further refine each free energy. }
    \label{fig:network}
  \end{center}
\end{figure}

By combining together the results of all independent Markovian sequences,
the outcome of our Monte Carlo sampling consists of an alchemical
\emph{network}. Each node of the network represents a
compound, and an edge connecting two nodes $A$ and $B$ corresponds to
an alchemical transformation that was sampled via an MD simulation,
see Fig.~\ref{fig:network}. Each edge is characterized by the
free-energy differences $\Delta G^{A\rightarrow B}_{\textup{i}}$ in
the three fixed environments, $i=\textup{W, I, M}$.  The network
representation was created with {\sc
NetworkX}~\cite{hagberg2008exploring} and visualized with {\sc
Gephi}~\cite{bastian2009gephi}.

For each environment, the net free-energy difference along any closed
cycle in the network must be zero (as shown in green in
Fig.~\ref{fig:network}), by virtue of a free energy being a state
function.  We thus enforced this thermodynamic condition to optimize
the set of free-energy differences calculated from MD simulations.  We
employed the algorithm proposed by Paton~\cite{paton1969algorithm} to
identify the cycle basis that spans the alchemical network, i.e., each
cycle in the network can be obtained as a sum of the $N_{C}$ basis
cycles.  We denote the MD free-energy differences involved in at least
one basis cycle by $\Delta G^{j}_i,~j=1,..,N_{G},~i=\textup{W, I, M}$,
while nodes connected to only a single edge cannot be taken into
account.  For each environment, we optimized the set of free energies
$\Delta \hat{G}^{j}_i$ by minimizing the loss function
\begin{equation}
  \label{eq:ObjFct}
  \mathcal{L}_i
  =  \sum_{j=1}^{N_G} (\Delta G^j_i - \Delta \hat{G}^j_i)^2 +
  \sum_{k=1}^{N_c} \omega \Big(\sum_{j \in k}
  (-1)^{s_{j,k}}\Delta \hat{G}^j_i\Big)^2.
\end{equation}
While the first term ensures that the optimized free-energy
differences $\Delta \hat{G}^{j}_{\textup{i}}$ remain close to the MD
simulation results, the second term ($\omega=10.0$) penalizes
deviations from zero for each thermodynamic cycle within a basis
cycle.  The exponent $s_{j,k}$ controls the sign of the
free-energy difference in the cycle, taking values of $0$ or $1$.  To
minimize the cost functions, we employed the
Broyden-Fletcher-Goldfarb-Shanno method
(BFGS)~\cite{avriel2003nonlinear} (see Figs.~S1 and S4 for trimers and
tetramers, respectively).

\subsection{Machine learning}

We use kernel ridge regression~\cite{rasmussen2004gaussian}, where the
prediction of target property $p({\bf x})$ for sample ${\bf x}$ is
expressed as a linear combination of kernel evaluations across the
training points ${\bf x}^*_i$
\begin{equation}
  p({\bf x}) = \sum_i \alpha_i K({\bf x}^*_i, {\bf x}).
\end{equation}
The kernel consists of a similarity measure between two samples
\begin{equation}
  K({\bf x}, {\bf x}') = \exp \left( - \frac{||{\bf x}-{\bf
      x}'||_1}{\sigma} \right),
\end{equation}
which corresponds to a Laplace kernel with a city-block metric (i.e.,
$L_1$-norm), and $\sigma$ is a hyperparameter.  The representation
${\bf x}$ corresponds to the vector of water/octanol partitioning free
energies of each bead---it is described more extensively in the
Results.  The optimization of
the weights $\alpha$ consists of solving for the samples in the
training with an additional regularization term $\lambda$: ${\bf
\alpha} = ({\bf K} + \lambda \mathrm{I})^{-1}{\bf p}$.  The confidence
of the prediction is estimated using the predictive variance
\begin{equation}
  \epsilon = {\bf K}^{**} - ({\bf K}^*)^T ({\bf K} + \lambda
  {\bf I})^{-1} {\bf K}^*,
\end{equation}
where ${\bf K}^{**}$ and ${\bf K}^*$ represent the kernel matrix of
training with training and training with test datasets,
respectively~\cite{rasmussen2004gaussian}.  The two hyperparameters
$\sigma$ and $\lambda$ were optimized by a grid search, yielding
$\sigma = 100$ and $\lambda = 10^{-4}$. Learning curves are shown in
Figs.~S2 and S5 for trimers and tetramers, respectively.

\section{Results}

We consider the insertion of a small molecule across a
single-component phospholipid membrane made of
1,2-dioleoyl-\emph{sn}-glycero-3-phosphocholine (DOPC) solvated in
water.  The insertion of a drug is monitored along the collective
variable, $z$, normal distance to the bilayer midplane
(Fig.~\ref{fig:intro}b).  We focus on three thermodynamic state points
of the small molecule: the bilayer midplane (``M''), the
membrane-water interface (``I''), and bulk water (``W'').  We link
these quantities in terms of transfer free energies, e.g., $\dGwm$
denotes the transfer free energy of the small molecule from water to
the bilayer midplane.

\subsection{Importance sampling}

We ran MC simulations across CG linear trimers and tetramers (results
for tetramers are shown in the SI), randomly changing a bead type,
calculating the relative free energy difference between old and new
compound in the three different environments, and accepting the trial
compound using a Metropolis criterion on the water/interface transfer
free energy $\dGwi$ (Fig.~\ref{fig:intro}a and Eq.~\ref{eq:metr_mc}).
This criterion aims at selecting compounds that favor partitioning at
the water/membrane interface.

The MC algorithm yielded an acceptance ratio of 0.2.  While initially
most trial compounds contributed to expand the database, the sampling
scheme quickly reached a stable regime where roughly half of the
compounds had already been previously visited.  Because each
free-energy calculation is expensive, we avoid recalculating identical
alchemical transformations to help efficiently converge the protocol.

To monitor and control the possible accumulation of statistical error
during the MC chain of alchemical transformations, we optimized the
network of sampled free energies---i.e., the one obtained
by combining together the results of all independent MC sequences---according to thermodynamic
constraints (Sec.~\ref{sec:topt}). A short MC sequence of accepted
compounds is shown in Fig.~\ref{fig:network}. We display the sequence
within the network of sampled compounds, each node being represented
by the set of Martini bead types involved. We find a large number of
closed paths within this network: since the free energy is a state
function, the closed path represents a thermodynamic cycle---it must
sum up to zero. We thus enforced this condition on the whole set of
basis cycle in the network and for each of the three environments to determine a set of optimized free-energy
differences, at the same time pushing the optimized values to remain
close to MD simulation results, see Eq.~\ref{eq:ObjFct}. We stress
that many free energies are involved in multiple basis cycles,
enhancing the robustness of the optimization by combining constraints.

The outcome of the optimization for both trimers and tetramers is
presented in the supporting information (Fig.~S1 and S4,
respectively). We found small modifications of the free energies
calculated via MD simulations to be sufficient to virtually enforce a
zero net free energy over the whole set of basis cycles.  Aggregate
changes along cycles never exceeded 0.2, 0.1, and 0.01~kcal/mol at the
interface, in water, and in the membrane core.

Any cycle in the network can be written as a combination of basis
cycles. As such, the condition enforced in the optimization offers
free-energy differences between two compounds to be calculated along
\emph{any} path connecting them in the network. This highlights the
robustness of our optimization scheme and hinders a significant
accumulation of errors in the relative free energies along a sequence
of compounds.

By combining together the results for the three thermodynamic environments through Eq.~\ref{eq:alch_transf},
we thereby obtain an \emph{optimized} network of compounds whose edges feature relative
transfer free energies $\dGwi$ and $\dGim$ between the two connecting
nodes.  Relative free energies can be summed up to trace the total
change in free energy during a path. This only leaves us to determine
the \emph{absolute} transfer free energies $\dGwi$ and $\dGim$ for an
arbitrary starting compound. These transfer free energies were
extracted from the potential of mean force, $G(z)$
(Fig.~\ref{fig:intro}b), calculated following the simulation protocol
described in Ref.~\onlinecite{menichetti2017silico}. A number of
reference PMFs for trimers and tetramers were calculated.  The lack of
compounding of the statistical errors, as evidenced by our
thermodynamic-cycle optimizations, and the sufficient number of MC
cycles make the choice of reference compounds insignificant.

\subsection{Machine learning}

The ML models used here infer the relationship between the
CG composition of a compound and its various transfer free energies. A
key component of an efficient ML model is its
representation~\cite{huang2016communication}. It should include
enough information to distinguish a compound's chemical composition
and geometry, as well as encode the physics relevant to the target
property~\cite{faber2018alchemical}. Because the CG compounds all
consist of beads arranged linearly and equidistant, we have found that
encoding the geometry had no benefit to the learning (data not shown).
Instead we simply encode the water/octanol partitioning of each bead,
yielding for linear trimers
$
{\bf x} = \left(\dGwol^{(1)}, \dGwol^{(2)}, \dGwol^{(3)}\right).
$
Reference values for $\dGwol^{(i)}$ were extracted from alchemical
transformations of each bead type between the two bulk
environments~\cite{Bereau2015}. Note that while the problem we
consider in this work contains reflection symmetry for the compounds
(i.e., $ABC$ is equivalent to $CBA$), we did not need to encode this
in the representation.  Instead we sorted the bead arrangement when
generating compounds for the importance sampling and machine learning.

When trained on most of
the MC-sampled data, we obtained out-of-sample mean absolute errors
(MAE) as low as 0.2~kcal/mol for $\dGwi$ and $\dGim$, on par with the
statistical error of the alchemical transformations (see Fig.~S2).
Remarkably, the prediction of $\dGwm$ converges to an MAE lower than
0.05~kcal/mol, illustrative of the strong correlation between
water/octanol and water/membrane free energies in
Martini~\cite{menichetti2017silico}. For all three quantities we
monitor a correlation coefficient above 97\%, indicating excellent
performance.

\begin{figure}[htbp]
  \begin{center}
    \includegraphics[width=\linewidth]{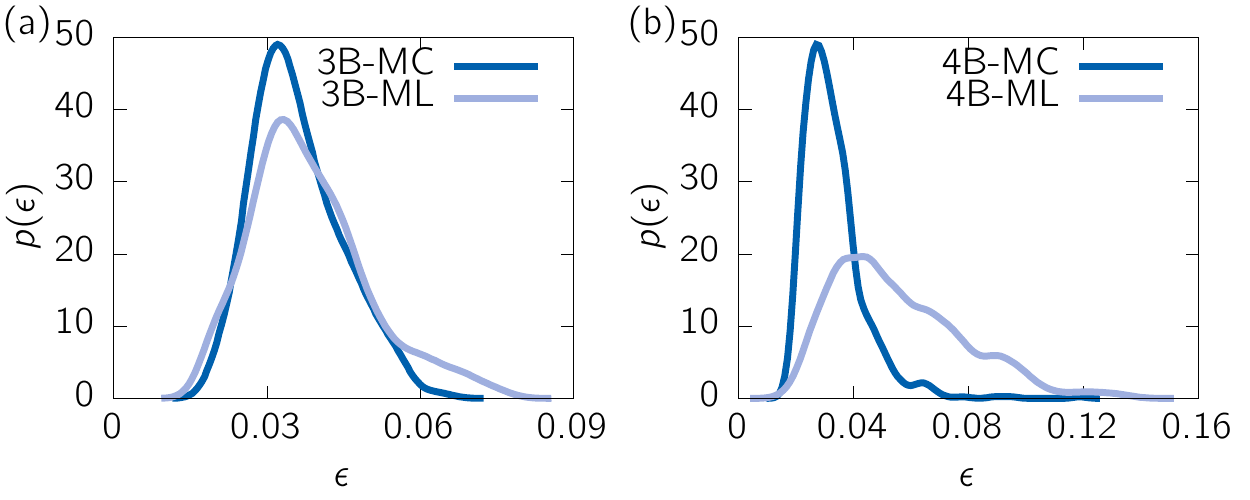}
    \caption{Distribution functions of confidence intervals, $p(\epsilon)$ (see Methods for definition), for both the out-of-sample predictions within the MC-sampled compounds and the ML-predicted compounds. (a) Trimers and (b) tetramers.}
    \label{fig:pred_var}
  \end{center}
\end{figure}

Next, we train our ML model on the entire dataset of MC-sampled
compounds.  We use this model to predict all other CG linear
trimers---a similar protocol was applied to tetramers.  Because of the
importance-sampling scheme, the predicted compounds will typically
feature different characteristics, e.g., more polar compounds that
would preferably stay in the aqueous phase.  As such the ML model is
technically extrapolating outside of the training set. As a measure of
homogeneity between training and validation sets,
Fig.~\ref{fig:pred_var} displays the distributions of confidence
intervals (see Methods) between out-of-sample predictions and the
expansion of the dataset. While we find significant overlap between
the MC and ML distributions for trimers, we observe larger deviations
in the case of tetramers.

\begin{figure}[htbp]
  \begin{center}
    \includegraphics[width=0.95\linewidth]{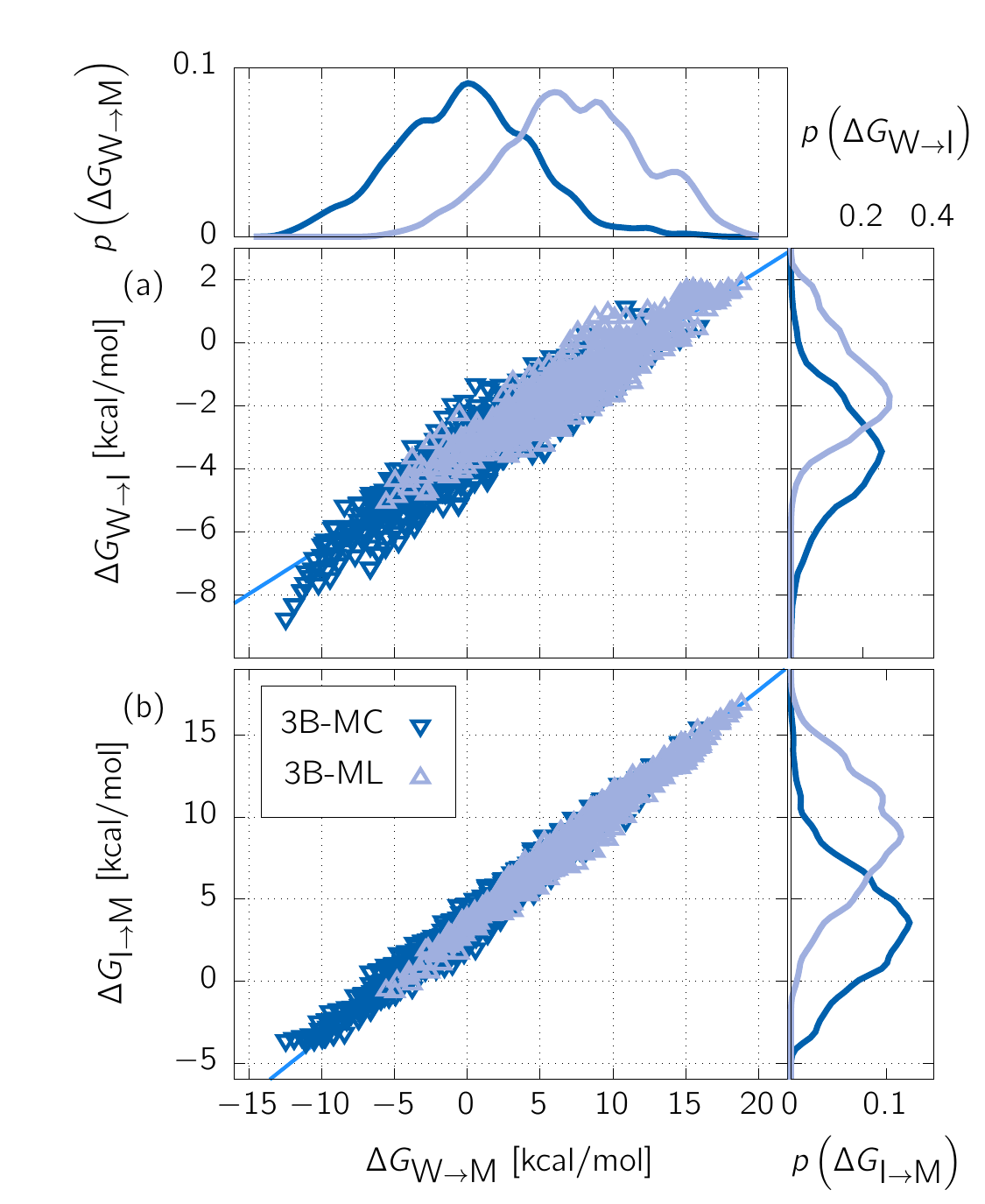}
    \caption{(a) Transfer free energies from water to interface
      $\dGwi$ as a function of the compound’s water/membrane
      partitioning free energy, $\dGwm$.  The dark and light blue
      points depict corresponding quantities for trimers estimated
      from MC sampling (3B-MC) and the ML predictions (3B-ML),
      respectively. Linear fits highlight the molecular-weight
      dependence.  (b) Transfer free energies from the interface to
      the membrane $\dGim$ as a function of the compound's
      water/membrane partitioning free energy, $\dGwm$.  The coverages
      are projected down along a single variable on the sides.  Error
      bars for 3B-MC are on par with the datapoint sizes (not shown).}
    \label{fig:linear}
  \end{center}
\end{figure}

The extrapolation can also be seen in the projections of predicted
transfer free energies, highlighting distinct coverages of sampled and
predicted trimer compounds (Fig.~\ref{fig:linear}).  However, the main
panels (a) and (b) display notable linear relations between transfer
free energies---similar behavior is found for tetramers (Fig.~S3).
Importantly, similar linear relations had already been observed for CG
unimers and dimers, highlighting thermodynamic relations for the
transfer between different effective bulk
environments~\cite{menichetti2017silico}.  We also argue that the ML
models do not simply learn linear features, since we optimize
independent models for the different predicted transfer free energies.
The linear behavior displayed across both sampled and predicted
compounds testifies to the robustness of the ML model, despite the
extrapolation.

\begin{figure*}[htbp]
  \begin{center}
    \includegraphics[width=0.8\linewidth]{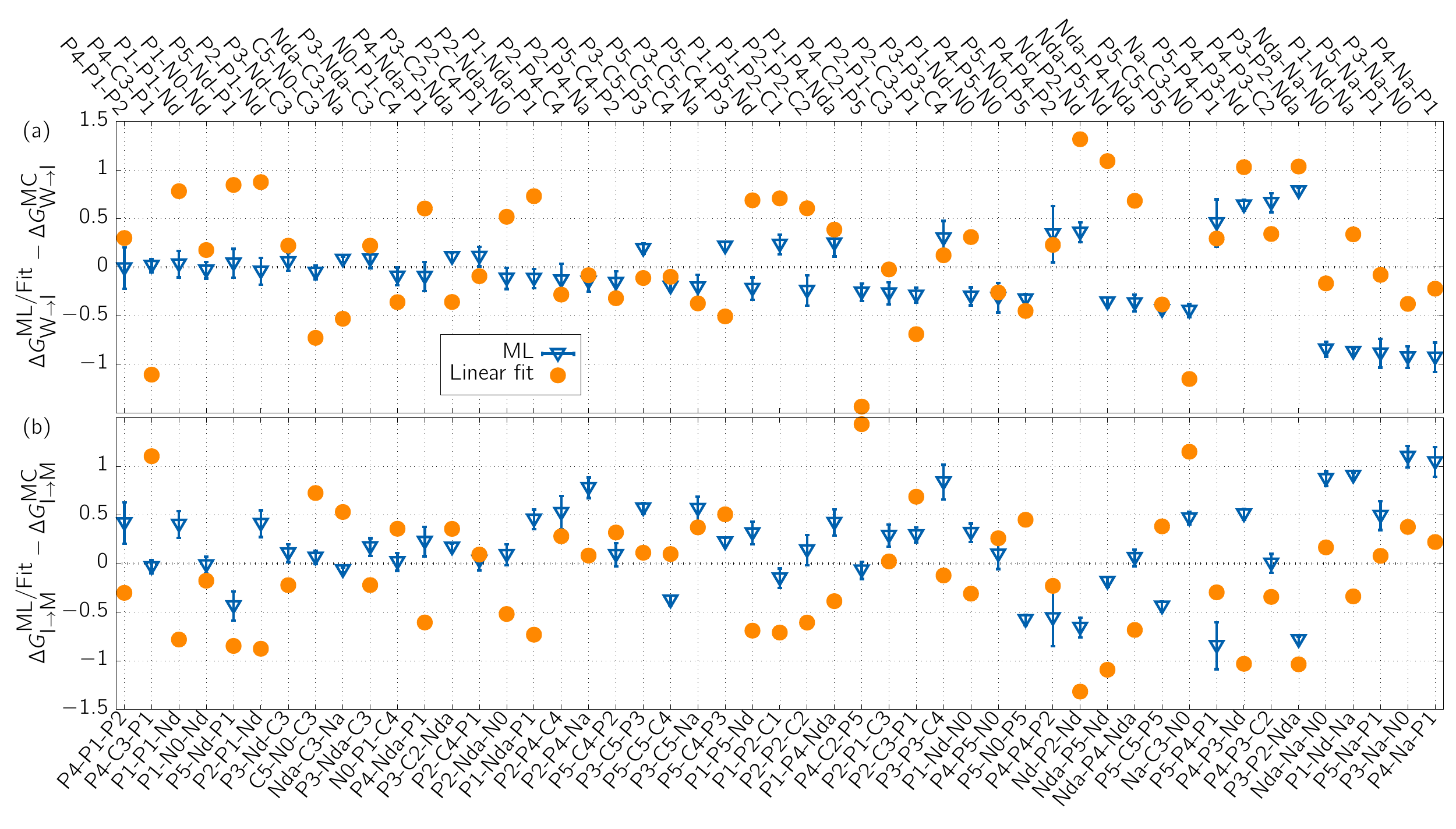}
     \caption{Deviations of the ML and linear-fit (``Fit'')
     predictions from reference alchemical transformations (``MC'')
     for (a) $\dGwi$ and (b) $\dGim$. Error bars for the ML model
     display the 95\% confidence intervals form the predictive
     variance. Compounds are sorted according to the ML prediction
     error for $\dGwi$.  Free energies displayed in units of
     kcal/mol.}
    \label{fig:linearFit_vs_ML}
  \end{center}
\end{figure*}

The ML predictions also offer higher accuracy compared to simple
linear fits: We selected a small set of 50 reference compounds
spanning the entire dataset and measured the performance of the ML
predictions and linear regression.  The deviation of both predictions
against reference alchemical transformations for each compound is
shown in Fig.~\ref{fig:linearFit_vs_ML}, displaying predictions for
$\dGwi$ and $\dGwm$.  We find a mean-absolute error (MAE) of 0.3 and
0.5~kcal/mol for the ML and linear fit, respectively.  The linear
regressions display equal but opposite errors between $\dGwi$ and
$\dGwm$, by construction. The compounds are sorted according to the ML
deviation of $\dGwi$. Interestingly, this ranking of compounds shows
no clear pattern: for instance, it only correlates weakly with
hydrophobicity (21\%).  On the other hand hydrophobicity correlates
much more strongly with the predictive variance (50\%).  While the
latter naturally stems from the representation, the absence of
correlation between ML deviation and hydrophobicity points at a more
complex structure of the interpolation space (i.e., the CG chemical
space)---a feature that will only worsen when learning atomistic
compounds.  Complementary information can be further probed from
Fig.~\ref{fig:linearFit_vs_ML} by comparing the ML deviations between
$\dGwi$ and $\dGim$: we observe a correlation coefficient of 63\%
between the unsigned prediction errors.  Training two independent ML
models on identical subsets of chemical space leads to large
correlations, further emphasizing the predominance of the
interpolation space over the target property when learning.

A systematic coarse-graining of compounds in the GDB~\cite{Fink2007}
using {\sc Auto-Martini}~\cite{Bereau2015} was performed to identify
small organic molecules that map to CG linear trimers. The algorithm
is deterministic, such that it leads to a unique mapping from molecule
to CG representation. We identified
1.36~million compounds, for which we can associate all three transfer
free energies, $\dGwm$, $\dGwi$, and $\dGim$.  We note that the
sampled and predicted CG representations amount to similar numbers of
compounds, such that the ML boosting introduced here offers an
additional 0.8~million compounds to the database.  The database is
provided as supporting material for further data analysis.

\section{Conclusions}

The overwhelming size of chemical space naturally calls for
statistical techniques to analyze it.  A variety of data-driven
methods such as quantitative structure-property relationships (QSPR)
and ML models at large have been applied to chemical
space~\cite{rupp2012fast, faber2016machine, bartok2017machine,
zhang2017machine}.  While sparse databases easily lead to
overfitting~\cite{swift2013back}, a dense coverage can offer
unprecedented insight~\cite{ramakrishnan2014quantum}.  Here we rely on
tools from statistical physics to ease the exploration of chemical
space: the application of importance sampling guides us toward the
subset of molecules that enhances a desired thermodynamic property.
This approach is similar to recent generative ML
models~\cite{sanchez2018inverse}, but without the a priori requirement
for labeled training data. 

In this work, we provide estimates of different transfer free energies (e.g.,
from water to the membrane interface, $\dGwi$) for a large number of
CG compounds.  The combination of alchemical transformations with MC
sampling motivates the calculation of free energies \emph{relative} to
the previous compound (e.g., $\Delta G^{A\rightarrow
B}_{\textup{I}}$). Estimating the stability of compound $C_i$, sampled
at the $i$-th step of an MC procedure, requires the summation of all
previous free-energy contributions, all the way to the initial
compound, for which we computed absolute free energies from umbrella
sampling.  Because each step in the MC procedure involves a free
energy, there is a statistical error that is compounded during this
reconstruction.  We can take advantage of thermodynamic cycles to
measure deviations from a net free energy of zero in a closed loop,
and thus estimate this compounding of errors. Remarkably, we find
deviations that are much smaller than the estimated statistical error
of the alchemical transformations.  This illustrates the robustness of
estimating free energies at high throughput using an MC scheme.

A conceptually-appealing strategy to expand the MC-sampled
distribution is through an ML model.  Effectively we train an ML model
on the MC samples and further boost the database with additional ML
predictions.  Unfortunately, the limited extrapolation behavior of
kernel models means that accurate predictions can only be made for
compounds \emph{similar} to the training set.  \emph{How} similar is
often difficult to estimate a priori.  Similarity metrics are often
based at the level of the ML's input space---here the molecular
representation.  In Fig.~\ref{fig:pred_var} we used the predictive
variance as a metric for the query sample's distance to the training
set~\cite{rasmussen2004gaussian}. 

Beyond similarity in the ML's input space via the predictive variance,
we also consider the target properties directly.  Our physical
understanding of the problem offers a clear requirement on the
transfer free energies, through the linear relationships shown in
Fig.~\ref{fig:linear}~\cite{menichetti2017silico}.  As such, the
thermodynamics of the system impose a physically-motivated constraint
on the predictions.  Rather than specific to each prediction, this
constraint is \emph{global} to the ensemble of data points. Satisfying
it grounds our predictions within the physics of the problem, ensuring
that we accurately expand the database.

Remarkably, we find that we can significantly expand our
database---doubling it for trimers and a factor of 10 for tetramers
(see SI)---while retaining accurate transfer free energies.  Unlike
conventional atomistic representations~\cite{faber2017prediction}, our
ML model is encoded using a CG representation, such that compounds
need only be similar at the CG level.  This CG similarity is strongly
compressed because ($i$) coarse-graining reduces the size of chemical
space~\cite{menichetti2017silico}, but also ($ii$) of a more
straightforward structure-property link~\cite{menichetti2018drug}. The
latter is embodied by the additive contribution of bulk partitioning
free energies for each bead, efficiently learning the molecular
transfer free energy in more complex environments.  All in all,
backmapping (Fig.~\ref{fig:intro}d) significantly amplifies the
additional region of chemical space reached by the ML model.  Our work
highlights appealing aspects of bridging physics-based methodologies
and coarse-grained modeling together with machine learning, offering
increased robustness and transferability to explore significantly broader
regions of chemical space.

\section{Supporting Information}

The attached supporting information contains additional details on the
optimization of thermodynamic cycles; the learning curves of the
machine learning model; and results on linear tetramers. In addition,
we provide databases for the transfer free energies of all trimers and
tetramers, as well as atomistic-resolution compounds that map to
trimers in a repository~\cite{zenodo}.

\section*{Acknowledgments}
The authors thank Alessia Centi and Clemens Rauer for critical reading
of the manuscript.  The authors acknowledge Chemaxon for an academic
research license of the Marvin Suite.  This work was supported by the
Emmy Noether program of the Deutsche Forschungsgemeinschaft (DFG) and
the John von Neumann Institute for Computing (NIC) through access to
the supercomputer JURECA at J\"ulich Supercomputing Centre (JSC).

\bibliography{biblio} 

\end{document}